\documentclass[
reprint,superscriptaddress,
amsmath,amssymb,aps,
]{revtex4-1}

\usepackage{graphicx}
\usepackage{currvita}
\usepackage{dcolumn}
\usepackage{bm}
\usepackage{color, soul}
\usepackage{amsmath}
\usepackage{natbib}
\usepackage{bm}
\usepackage{times}
\usepackage{url}
\usepackage{amssymb}
\usepackage{braket}
\usepackage{mathtools}
\usepackage{commath}
\usepackage{bigints}
\usepackage{booktabs}
\usepackage{float}
\usepackage{multirow}
\usepackage{graphicx}
\usepackage{dcolumn}
\usepackage{bm}
\usepackage{hyperref}
\usepackage[mathlines]{lineno}
\begin{document}
	
\preprint{APS/123-QED}
	
\title{Magnetic and structural properties of Ni-substituted  magnetoelectric Co$_{4}$Nb$_2$O$_9$  }

\author{Hadi Papi}
\affiliation{Department of Physics, Isfahan University of Technology, Isfahan 84156-83111, Iran}
\affiliation{Laboratory for Quantum Magnetism, Institute of Physics, Ecole Polytechnique Féderale de Lausanne, CH-1015 Lausanne, Switzerland}
\author{Virgile Yves Favre}
\affiliation{Laboratory for Quantum Magnetism, Institute of Physics, Ecole Polytechnique Féderale de Lausanne, CH-1015 Lausanne, Switzerland}
\author{Hossein Ahmadvand}
\email{ahmadvand@cc.iut.ac.ir}
\affiliation{Department of Physics, Isfahan University of Technology, Isfahan 84156-83111, Iran}
\author{Mojtaba Alaei}
\affiliation{Department of Physics, Isfahan University of Technology, Isfahan 84156-83111, Iran}
\author{Mohammad Khondabi}
\affiliation{Department of Physics, Isfahan University of Technology, Isfahan 84156-83111, Iran}
\author{Denis Sheptyakov}
\affiliation{Laboratory for Neutron Scattering and Imaging, Paul Scherrer Institut, CH-5232 Villigen PSI, Switzerland}
\author{Lukas Keller}
\affiliation{Laboratory for Neutron Scattering and Imaging, Paul Scherrer Institut, CH-5232 Villigen PSI, Switzerland}
\author{Parviz Kameli}
\affiliation{Department of Physics, Isfahan University of Technology, Isfahan 84156-83111, Iran}
\author{Ivica \v{Z}ivkovi\'{c}}
\affiliation{Laboratory for Quantum Magnetism, Institute of Physics, Ecole Polytechnique Féderale de Lausanne, CH-1015 Lausanne, Switzerland}
\author{Henrik M. R\o{}nnow}
\email{henrik.ronnow@epfl.ch}
\affiliation{Laboratory for Quantum Magnetism, Institute of Physics, Ecole Polytechnique Féderale de Lausanne, CH-1015 Lausanne, Switzerland}	

	\date{\today}
	
	\begin{abstract}
The magnetic and structural properties of polycrystalline Co$_{4-x}$Ni$_x$Nb$_2$O$_9$ (x=1,2) have been investigated by neutron powder diffraction, magnetization and heat capacity measurements, and density functional theory (DFT) calculations. For $x$=1, the compound crystallizes in the trigonal P$\bar{\textrm{3}} $c1 space group. Below T$_N$=31 K it develops a weakly non-collinear antiferromagnetig structure with magnetic moments in the $ab$-plane. The compound with $x$=2 has crystal structure of the orthorhombic Pbcn space group and shows a hard ferrimagnetic behavior below T$_C$=47 K. For this compound a weakly non-collinear ferrimagnetic structure with two possible configurations in $ab$ plane was derived from ND study. By calculating magnetic anisotropy energy via DFT, the ground-state magnetic configuration was determined for this compound. The heat capacity study in magnetic fields up to 140 kOe provide further information on the magnetic structure of the compounds.

	\end{abstract}
	
	\maketitle
	
\section{INTRODUCTION}\label{Int}

Magnetoelectric materials which provide electric (magnetic) field manipulation of magnetization (polarization) have attracted significant attention due to both their interesting fundamental physics and potential applications \cite{Spaldin, Feibeg, SHUAI, Fennie, Yurong, Soda,Kimura}. Their potential application in data storage, for instance, lies in the feasibility of controlling the magnetic information by applying electric field \cite{Eerenstein,Maignan}.  
    
In a class of magnetoelectric materials such as Cr$_{2}$O$_{3}$, application of a magnetic field induces electric polarization below their magnetic ordering temperature. This induced electric polarization is zero in the absence of magnetic field and increases linearly with applied field \cite{iyama}. The family of M$_{4}$A$_{2}$O$_{9}$ (M: Co, Mn and A: Nb, Tb) compounds was initially reported by Fischer et al.  \cite{FISCHER} to show a similar magnetoelectric coupling under application of magnetic field below their magnetic ordering temperature. In recent years, several works have been published considering the magnetic, structural and magnetoelectric features in this series of compound \cite{DengPRB,FengSCI,Kolodiazhnyi,Maignan,Khanh1,KhanhPRB,DengPRB,Solovyev,Narayanan,Xie,Yanagi,Liu}. Among these compounds, Co$_{4}$Nb$_{2}$O$_{9}$ has been reported to show high magnetoelectric coupling at the vicinity of its Neel transition temperature \cite{Kolodiazhnyi,Solovyev,FengSCI}. It crystallizes in the trigonal P$\bar{\textrm{3}} $c1 space group (No. 165) and shows an antiferromagnetic (AFM) phase transition at around 27 K. Different magnetic structures including collinear AFM structure with moments alignment along the c axis \cite{BERTAUT}, collinear structure with moments lying in ab plane and canting toward the c axis \cite{KhanhPRB}, and more recently an in-plane non-collinear magnetic configuration \cite{DengPRB} have been reported for Co$_{4}$Nb$_{2}$O$_{9}$ from neutron diffraction analysis.

In order to gain more insights into the magnetic structure of this compound, it is helpful to substitute Co$^{2+}$ by magnetic ions. Recently, it was shown that in the Mn$^{2+}$ doped compound, Co$_{4-x}$Mn$_x$Nb$_2$O$_9$, the noncollinear AFM structure is stable up to x=3.9, which is due to the strong easy-plane anisotropy of Co$^{2+}$ \cite{Mn-Dop}. 
Here, we study the effects of substitution of Co$^{2+} $ by Ni$^{2+} $ on magnetic properties of Co$_{4-x}$Ni$_x$Nb$_2$O$_9$ by means of neutron diffraction (ND), magnetization and heat capacity measurements, and density functional theory (DFT) calculations. For x=1, the compound crystallizes in the P$\bar{\textrm{3}} $c1 space group with an in-plane weakly noncollinear AFM configuration, in agreement with that recently reported for Co$_{4}$Nb$_{2}$O$_{9}$ \cite{DengPRB}. While the compound with x=2, similar to Ni$_4$Nb$_2$O$_9$ \cite{Ehrenberg}, has crystal structure of the orthorhombic Pbcn space group. A weakly non-collinear ferrimagnetic structure with moments lying along the \textbf{b} axis is revealed for this compound. 
\section{EXPERIMENTAL DETAILS}\label{Exp}
Polycrystalline samples of Co$_4$Nb$_2$O$_9$ (CN0), Co$_3$Ni$_1$Nb$_2$O$_9$ (CN1) and Co$_2$Ni$_2$Nb$_2$O$_9$ (CN2) were synthesized using solid-state reaction method. Stoichiometric mixture of high purity Co$_3$O$_4$, NiO and Nb$_2$O$_5$ were thoroughly ground by hand in agate mortar, and annealed in air at 1373 K for 8 h. The obtained powders were again ground, pressed into pellets and sintered at 1473 K for 30 h. Samples were initially examined using x-ray diffraction. No secondary phases were detected in CN0 and CN1, however, about 5\% phase of NiNb$_2$O$_6$ contributes to CN2 diffraction pattern.

Neutron diffraction (ND) measurements were performed at the Swiss Spallation Neutron Source (SINQ), Paul Scherrer Institute. Powder samples were loaded into 6-mm-diameter vanadium cans. In order to identify the crystal structure parameters, ND data were taken on the High-Resolution Powder Diffractometer for Thermal Neutrons (HRPT) \cite{HRPT} at 1.6 and 300 K using two wavelengths of 2.45 and 1.494 \AA. For magnetic structure investigations, ND data were collected on the Cold Neutron Powder Diffractometer (DMC). A wavelengths of 2.458 \AA. was used and data were collected in the 1.5 - 45 K and 1.5 - 60 K ranges for CN1 and CN2, respectively. 

The ND patterns were analyzed by Rietveld refinement \cite{rietveld} using the FULLPROF program suite \cite{fullprof}. In order to determine the possible magnetic configurations, irreducible representation analysis was done using SARAh \cite{sarah}. The visualization software VESTA \cite{vesta} was used for displaying the crystal structure.

Magnetization and heat capacity experiments were conducted on pressed pellets. Magnetization as functions of temperature and applied magnetic field was measured by using a Superconducting Quantum Interference Device (SQUID) magnetometer. Heat capacity was measured using
a Quantum Design Physical Properties
Measurement System (PPMS).

\section{Computational Details}
We used the supercell program~\cite{supercell} to generate symmetrically independent atomic combination 
of Co and Ni. Then, to select the most stable atomic combination based on the total energy criterion, 
we employed DFT by using Quantum-Espresso~\cite{QE}. 
For exploring magetic easy axis, we used Fleur code~\cite{fleur} which is also based on DFT.
To improve Coulomb d-orbital interactions,
we used Hubbard $U$ correction to DFT (DFT+$\textrm{U}$) in magnetic anisotropy calculations.
In all of our DFT calculations, we estimated electronic exchange correlation energy by PBE functional~\cite{PBE}.

\section{RESULTS AND DISCUSSION}\label{RESULT}
\subsection{Neutron Diffraction: Nuclear and magnetic structure}\label{ND}
\begin{figure}[H]
	\begin{center}
		\includegraphics[width=9cm,angle=0]{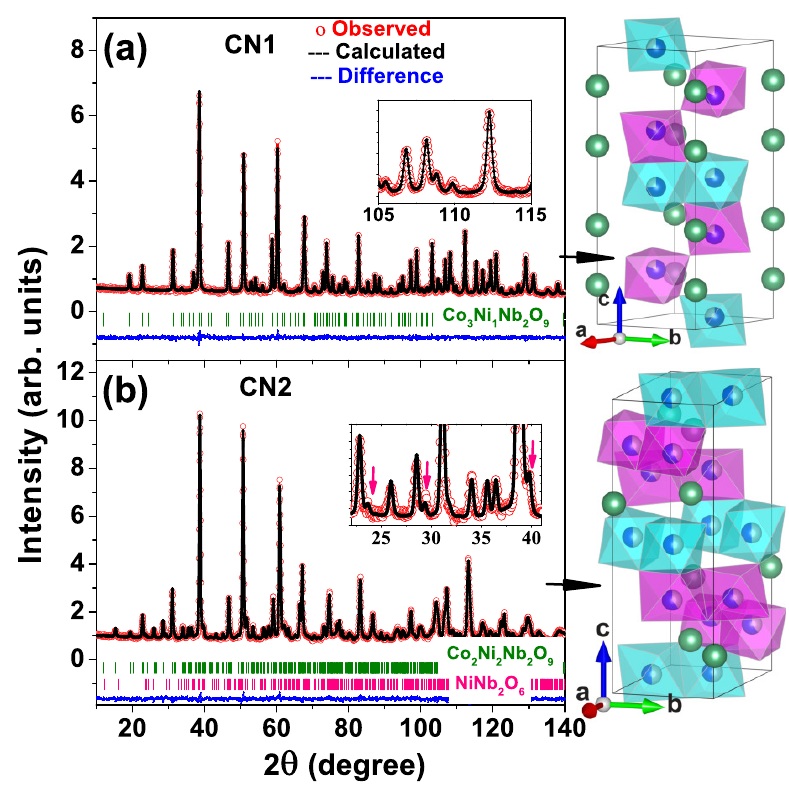}
		\caption{ND pattern obtained at room temperature using a wavelength of $\lambda = 1.49 Å$ on HRPT and associated crystal structure of (a)  Co$_3$Ni$_1$Nb$_2$O$_9$ (CN1) and (b) Co$_2$Ni$_2$Nb$_2$O$_9$ (CN2). The observation, calculation, and their difference are plotted in red circles, black and blue
			lines, respectively. The vertical bars in green mark the Bragg positions for the main phases and the pink ones mark the Bragg positions for secondary phase in CN2 case. Insets show zoomed view of the Rietveld fits. Red arrows in inset of (b) mark peaks related to the secondary phase. In crystal structures different crystallographic sites shared by cobalt and nickel are surrounded by octahedrons with different colors.     
			\label{crys}}
	\end{center}
\end{figure}

\begin{figure}[H]
	\begin{center}
		\includegraphics[width=8.5cm,angle=0]{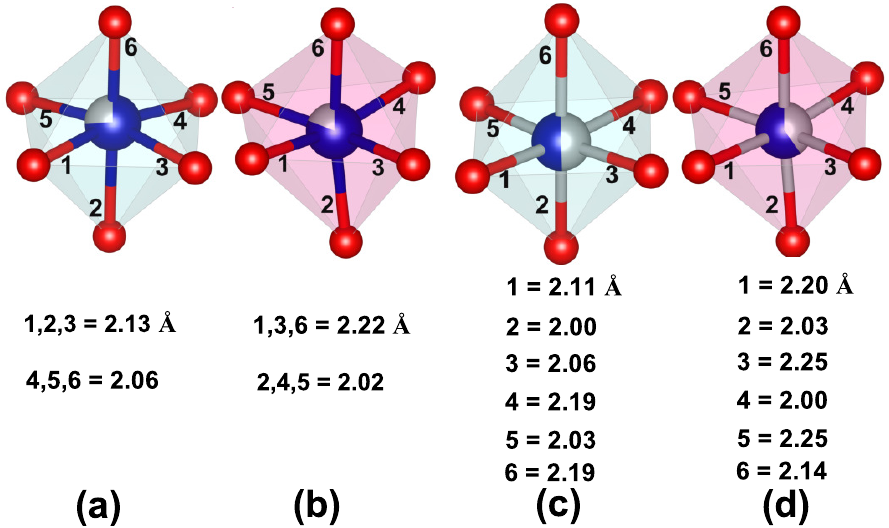}
		\caption{Octahedral environment of crystallographic sites occupied by Co/Ni (a) site1 of  Co$_3$Ni$_1$Nb$_2$O$_9$ (CN1), (b) site2 of CN1, (c) site1 of Co$_2$Ni$_2$Nb$_2$O$_9$ (CN2) and (d) site2 of CN2.              
			\label{oct}}
	\end{center}
\end{figure}

The Rietveld refinement of the ND data of CN1 and CN2 samples collected with HRPT at room temperature and their corresponding crystal structures are shown in Fig. \ref{crys}. The ND pattern shows that CN1, like CN0, crystallizes in the P$\bar{\textrm{3}} $c1 space group. There are two non-equivalent crystallographic sites for magnetic atoms shared by Co and Ni, namely (Co/Ni)1 and (Co/Ni)2. There are also two distinct sites for oxygen atoms O1 and O2. (Co/Ni)1 is surrounded by six O2 atoms forming a distorted octahedron. (Co/Ni)2 is also in the center of another distorted octahedron containing three O1 and three O2 atoms. Both octahedrons include two different (Co/Ni)-O bond lengths (Fig. \ref{oct} (a,b)). On the other hand, CN2, similar to Ni$_4$Nb$_2$O$_9$ \cite{Ehrenberg}, has crystal structure of the orthorhombic Pbcn space group. It has two individual sites for Co/Ni, and five crystallographically different oxygen sites. Comparing to CN1, in CN2 (Co/Ni)1 and (Co/Ni)2 form more distorted octahedrons with six different Co/Ni-O distances(Fig. \ref{oct} (c,d)).   
The refinement results for nuclear structures of both samples are summarized in Table \ref{tab1}. The refined values of occupation for the shared sites give a stoichiometry of Co$_{3.1}$Ni$_{0.9}$Nb$_{2}$O$_{9} $ and Co$_{2.12}$Ni$_{1.88}$Nb$_{2}$O$_{9} $ for the CN1 and CN2 samples, respectively. 

\begin{table*}[!htbp]
	\begin{center}
		\caption{ Crystal structure information of  Co$_3$Ni$_1$Nb$_2$O$_9$ (CN1) and Co$_2$Ni$_2$Nb$_2$O$_9$ (CN2) samples derived  from Rietveld refinement. The neutron diffraction data were collected at room temperature using a wavelength of $\lambda = 1.49 \AA$ on HRPT. }\label{tab1}

		\begin{tabular}{cccccccc}

			\hline
			\hline
			Sample&Space Group&$a[\AA]$&$b[\AA]$ &$c[\AA]$&$\alpha$& $\beta$& $\gamma$  \\
			CN1&$P \bar{3}c 1$ & 5.14958(1)    &	5.14958(1) &14.10625(2)&90& 90 & 120  \\
			CN2&$ P b c n $&8.81177(3)&5.11453(2)  &14.31992(5)&90&	90&	90  \\
			\hline
			&Atom&Wyck Symb&$x$&$y$ &$z$& Biaso$[\AA^{2}]$ &Occupancy \\
			\hline
			\multirow{5}{*}{CN1}&Nb &$4c$& 0&0 &0.1424(2)&0.54(5)& 1 \\
			&Co1$ / $ Ni1&$4d$& $\frac{\textrm{1}}{\textrm{3}}$ &$\frac{\textrm{2}}{\textrm{3}}$ &0.0131(4)&0.46(7)& 0.74(1)$ / $0.26(1)   \\
			&Co2$ / $Ni2 &$4d$&    $\frac{\textrm{1}}{\textrm{3}}$ & 	$\frac{\textrm{2}}{\textrm{3}}$ & 	0.3099(5)  &	0.46(7)	&0.81(1)$ / $0.19(1)\\
			&O1  &$6f$&  0.2882(5)  &	0  &0.25 &	0.65(4)&1\\
			&O2 &$12g$& 0.3425(2)  &	0.3203(5)  &	0.0837(2)  &	0.72(4)	&1\\
			\hline
			\multirow{8}{*}{CN2}&Nb&$ 8d $ &  0.0225(2) &-0.002(1)&0.3557(1)  &0.50(4)& 1  \\
			&Co1$ / $Ni1&$ 8d $&0.1646(5)  &0.508(1)  &0  &0.66(4)& 0.50(1)$ / $0.50 (1)  \\
			&Co2$ / $Ni2&$ 8d $&0.3355(6)   &-0.002(1)&0.1895(2)  &0.66(4)  & 0.60(1)2$ / $0.40(1)  \\ 

			&O1  &$ 4c $&0  &0.288(2)  &0.25  &0.83(11)& 1   \\
			&O2 &$ 8d $&0.1646(5) &0.166(1)  &0.4237(2)  &0.97(8)  & 1  \\
			&O3  &$ 8d $&0.1657(7)  &0.172(1)  &0.0953(3)&0.90(7)  &1   \\
			&O4&$ 8d $&0.3526(5)  &0.352(1)  &0.2520(3)  &0.51(6)& 1  \\
			&O5 &$ 8d $&0.4957(5)  &0.168(1)  &0.0838(4)  &0.75(8)& 1   \\
			\hline

			&&&\multicolumn{4}{c}{Reliability indexes of the refinements }\\
			\cline{3-8}
			&&\multicolumn{2}{c}{Bragg R-factor}&\multicolumn{2}{c}{RF-factor}&\multicolumn{2}{c}{$\chi^2$}\\
			CN1&&\multicolumn{2}{c}{2.26}&\multicolumn{2}{c}{1.85}&\multicolumn{2}{c}{1.19}\\
			CN2&&\multicolumn{2}{c}{2.59}&\multicolumn{2}{c}{1.86}&\multicolumn{2}{c}{1.49}\\
			\hline
			\hline
		\end{tabular}  		
		
	\end{center}

\end{table*}

\begin{figure*}[!]
	\begin{center}
		\includegraphics[width=18cm,angle=0]{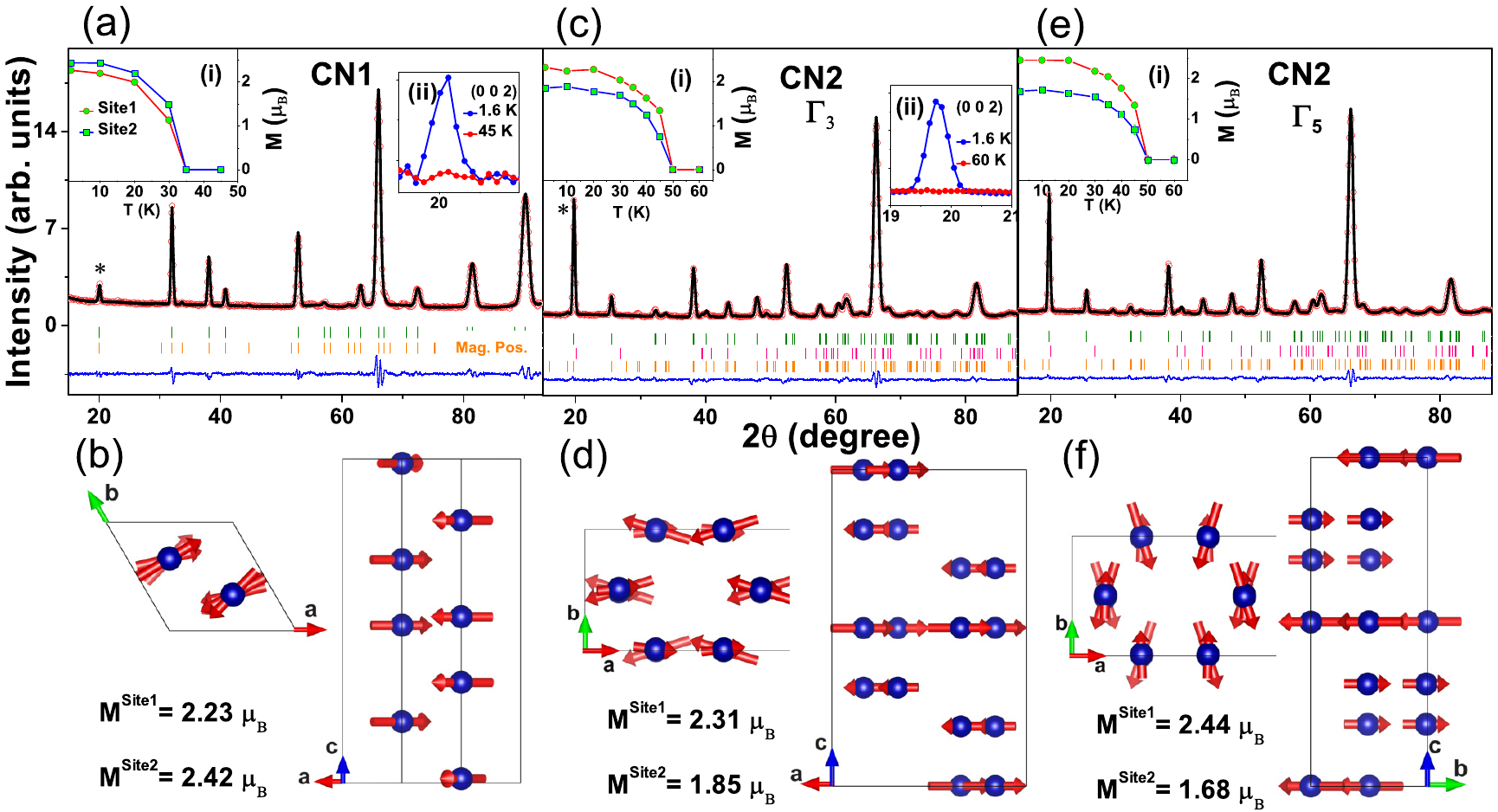}
		\caption{(a) Neutron diffraction pattern obtained at 1.7 K using a wavelength of $\lambda = 2.458 Å$ on DMC for  Co$_3$Ni$_1$Nb$_2$O$_9$ (CN1);  (b) magnetic structure of CN1; (c) ND pattern of Co$_2$Ni$_2$Nb$_2$O$_9$ (CN2) at 1.7 K, fitted based on $ \Gamma_{3} $ IR; (d) magnetic structure of CN2 related to $ \Gamma_{3} $; (e) ND pattern of CN2 at 1.7 K, fitted based on $ \Gamma_{5} $ IR; (f) magnetic structure of CN2 related to $ \Gamma_{5} $. Insets (i) are the temperature dependence of derived magnetic moments for site1 and site2 (Lines are guides to the eyes.); Insets (ii) show the peak profile for the main magnetic peak marked by a star at selected temperatures. In panels (a), (c) and (e), the observation, calculation, and their difference are plotted in red circles, black and blue lines, respectively. The vertical bars in green mark the nuclear Bragg positions for the main phases, the pink ones mark the Bragg positions for secondary phase in CN2, and the orange ones mark the magnetic Bragg positions.
			\label{dmc}}
	\end{center}
\end{figure*}

In order to probe magnetic structure of CN1 and CN2 compounds, we collected ND data on DMC in the temperature ranges of 1.6 - 45 K for CN1 and 1.6 - 60 K for CN2. For both cases magnetic peaks appear on top of nuclear ones showing that the propagation vector of the magnetic phase is \textbf{k}=(0,0,0). According to symmetry analysis for the space group P$\bar{\textrm{3}} $c1 and propagation vector \textbf{k}=(0,0,0), the magnetic structure of CN1 can be described by six candidate irreducible representations (IRs), namely $\Gamma_{1}-\Gamma_{6}$. Regarding all of these six potential IRs, it is found that only $ \Gamma_{6} $ (see Table \ref{IR}) provides the solution for magnetic structure of CN1. It fits the observed ND data with magnetic R-factor of 6.3. It leads to a weakly non-collinear AFM structure (Fig. \ref{dmc} (c)). For CN2 with the space group Pbcn and propagation vector \textbf{k}=(0,0,0), symmetry analysis proposes eight candidate IRs, $\Gamma_{1}-\Gamma_{8}$. Either $ \Gamma_{3} $ and $ \Gamma_{5} $ (see Table \ref{IR}) out of these eight IRs provide the solution for magnetic structure of CN2, however, the $ \Gamma_{5} $ model yields slightly better agreement factors of refinements. The magnetic R-factor is 6.88 for $ \Gamma_{3} $ and 4.18 for $ \Gamma_{5} $. Both of these two models suggest a weakly non-collinear ferrimagnetic structure for CN2. As it can be seen from Fig. \ref{dmc}(g) and (h) corresponding to $ \Gamma_{3} $ and $ \Gamma_{5} $ , the in plane magnetic moments have orientation dominantly along $ a$ and $ b $, respectively. In next section we will identify the most possible configuration between these two magnetic structures using DFT simulation. The complete decompositions of the magnetic representation is given in Tables SI and SII of of the supplementary materials \cite{supplementary}. The derived magnetic moments depend weakly on magnetic form factors provided for magnetic ions during the refinement. Since we have sites shared by Co and Ni, we refined once the data using Co$^{2+}$ magnetic form factor (MCO2) and once using Ni$^{2+}$ magnetic form factor (MNI2). As MCO2 and MNI2 are almost equal, the refined moments are very close to each other. For instance, the derived moments on site1 of CN2 are 2.45  and 2.43 $ \mu_B $ for MCO2 and MNI2, respectively. In next parts of this paper we deal with average magnetic moments refined based on MCO2 and MNI2. Derived magnetic moments at base temperature are given in Fig. \ref{dmc}. Insets (i) of Fig. \ref{dmc} show the temperature dependence of magnetic moments derived from the refinements.

\begin{table*}[!htbp]
	\begin{center}
		\caption{Basis vectors of irreducible representations $ \Gamma_{6} $ for the space group P$\bar{\textrm{3}} $c1, and of the $ \Gamma_{3} $ and $ \Gamma_{5} $ for the space group Pbcn, in both cases with propagation vector (0,0,0), for the magnetic Co/Ni sites (1/3,2/3,z) in P$\bar{\textrm{3}} $c1 and (x,y,z) in Pbcn, providing the solution for magnetic structures of the compounds Co$_3$Ni$_1$Nb$_2$O$_9$ (CN1) and Co$_2$Ni$_2$Nb$_2$O$_9$ (CN2), correspondingly. The complete decompositions of the magnetic representation is given in Tables SI and SII of the supplementary materials \cite{supplementary}. }\label{IR}
		\begin{tabular}{cccccccccc}
			\hline
			\hline

	    	&Co/Ni site&\multicolumn{2}{c}{($\frac{\textrm{1}}{\textrm{3}}$ ,$\frac{\textrm{2}}{\textrm{3}}$,z)}&\multicolumn{2}{c}{($\frac{\textrm{2}}{\textrm{3}}$ ,$\frac{\textrm{1}}{\textrm{3}}$,$\frac{\textrm{1}}{\textrm{2}}\textrm{-z}$)}&\multicolumn{2}{c}{($\frac{\textrm{2}}{\textrm{3}}$ ,$\frac{\textrm{1}}{\textrm{3}}$,$1$-z)}&\multicolumn{2}{c}{($\frac{\textrm{2}}{\textrm{3}}$ ,$\frac{\textrm{1}}{\textrm{3}}$,$\frac{\textrm{1}}{\textrm{2}}\textrm{+z}$)}\\
						\cline{3-10}

			\multirow{4}{*}{CN1}&$\Gamma_{6}$&\multicolumn{2}{c}{($\frac{\textrm{3}}{\textrm{2}}$,0,0)}&\multicolumn{2}{c}{($-\frac{\textrm{3}}{\textrm{2}}$,$-\frac{\textrm{3}}{\textrm{2}}$,0)}&\multicolumn{2}{c}{($-\frac{\textrm{3}}{\textrm{2}}$,0,0)}&\multicolumn{2}{c}{($\frac{\textrm{3}}{\textrm{2}}$,$\frac{\textrm{3}}{\textrm{2}}$,0)}\\
			&&\multicolumn{2}{c}{(0,$\frac{\textrm{3}}{\textrm{2}}$,0)}&\multicolumn{2}{c}{(0,$\frac{\textrm{3}}{\textrm{2}}$,0)}&\multicolumn{2}{c}{(0,$-\frac{\textrm{3}}{\textrm{2}}$,0)}&\multicolumn{2}{c}{(0,$-\frac{\textrm{3}}{\textrm{2}}$,0)}\\
			&&\multicolumn{2}{c}{($\frac{\sqrt{\textrm{3}}}{\textrm{2}}$,$\sqrt{\textrm{3}}$,0)}&\multicolumn{2}{c}{($-\frac{\sqrt{\textrm{3}}}{\textrm{2}}$,$\frac{\sqrt{\textrm{3}}}{\textrm{2}}$,0)}&\multicolumn{2}{c}{($-\frac{\sqrt{\textrm{3}}}{\textrm{2}}$,$-\sqrt{\textrm{3}}$,0)}&\multicolumn{2}{c}{($\frac{\sqrt{\textrm{3}}}{\textrm{2}}$,$-\frac{\sqrt{\textrm{3}}}{\textrm{2}}$,0)}\\
			&&\multicolumn{2}{c}{($-\sqrt{\textrm{3}}$,$-\frac{\sqrt{\textrm{3}}}{\textrm{2}}$,0)}&\multicolumn{2}{c}{($\sqrt{\textrm{3}}$,$\frac{\sqrt{\textrm{3}}}{\textrm{2}}$,0)}&\multicolumn{2}{c}{($\sqrt{\textrm{3}}$,$\frac{\sqrt{\textrm{3}}}{\textrm{2}}$,0)}&\multicolumn{2}{c}{($-\sqrt{\textrm{3}}$,$-\frac{\sqrt{\textrm{3}}}{\textrm{2}}$,0)}\\

			\hline
			&Co/Ni site &(x,y,z)&($\frac{\textrm{1}}{\textrm{2}}\textrm{-x}$,$\frac{\textrm{3}}{\textrm{2}}\textrm{-y}$,$\frac{\textrm{1}}{\textrm{2}}\textrm{+z}$)&($1\textrm{-x}$,y,$\frac{\textrm{1}}{\textrm{2}}\textrm{-z}$)&($\frac{\textrm{1}}{\textrm{2}}\textrm{+x}$,$\frac{\textrm{3}}{\textrm{2}}\textrm{-y}$,$1\textrm{-z}$)&($1\textrm{-x}$,$1\textrm{-y}$,$1\textrm{-z}$)&($\frac{\textrm{1}}{\textrm{2}}\textrm{+x}$,$\textrm{-}\frac{\textrm{1}}{\textrm{2}}\textrm{+y}$,$\frac{\textrm{1}}{\textrm{2}}\textrm{-z}$)&(x,$1\textrm{-y}$,$\frac{\textrm{1}}{\textrm{2}}\textrm{+z}$)&($\frac{\textrm{1}}{\textrm{2}}\textrm{-x}$,$\textrm{-}\frac{\textrm{1}}{\textrm{2}}\textrm{+y}$,z)\\
			\cline{3-10}			
			\multirow{7}{*}{CN2}&$ \Gamma_{3} $&(100)&(100) &(100) &(100)&(100)&(100) &(100)&(100)\\
			&&(010)&(0-10)&(0-10)&(010)&(010)&(0-10)&(0-10)&(010)\\
			&&(001)&(00-1)&(001)&(00-1)&(001)&(00-1)&(001)&(00-1)\\

			&$ \Gamma_{5} $&(100)&(-100)&(-100)&(100)&(100)&(-100)&(-100)&(100)\\
			&&(010)&(010)&(010)&(010)&(010)&(010)&(010)&(010)\\
			&&(001)&(001)&(00-1)&(00-1)&(001)&(001)&(00-1)&(00-1)\\
			
			\hline
			\hline
		\end{tabular}

	\end{center}
\end{table*}

\subsection{Theoretical predictions}
For the most of theoretical calculation one needs to know the exact atomic configuration of Co and Ni in CN1 and CN2. 
For CN1, there are 4 independent atomic combinations among 16 possible combinations 
(4 possible substitutions for Ni$_1$ atom (Ni$_2$) in the first (second) 4d Wyckoff position, see Table \ref{tab1}).
For CN2, in the case of equal occupancy of Co2/Ni2, there are 644 symmetrically independent atomic combinations among 4900 possible combinations
($\frac{8!}{4!4!}$ possible substitutions for 4 Ni$_1$ (Ni$_2$) atoms in the second (third) 8d Wyckoff position, see Table \ref{tab1}).
According to Table \ref{tab1},  the occupancy of 8d Wyckoff position for Co$_2$/Ni$_2$  is 0.6/0.4. 
	If we want to simulate the exact occupancy of 0.6/0.4, we need a supercell which makes the DFT calculations very time-consuming. 
	Fortunately for 0.625/0.375 occupancy of Co$_2$/Ni$_2$, which is a good approximation for 0.6/0.4, we don't need to use supercell.
	In this case, there are 490 symmetrically independent atomic combinations among 3920 possible combinations
	($\frac{8!}{4!4!}$ possible substitutions for 4 Ni$_1$ atoms in the second 8d Wyckoff position and $\frac{8!}{3!5!}$ 
	possible substitutions for 3 Ni$_2$ atoms in the third 8d Wyckoff position, see Table \ref{tab1}).

\begin{figure}[!]
	\begin{center}
		\includegraphics[width=8.5cm,angle=0]{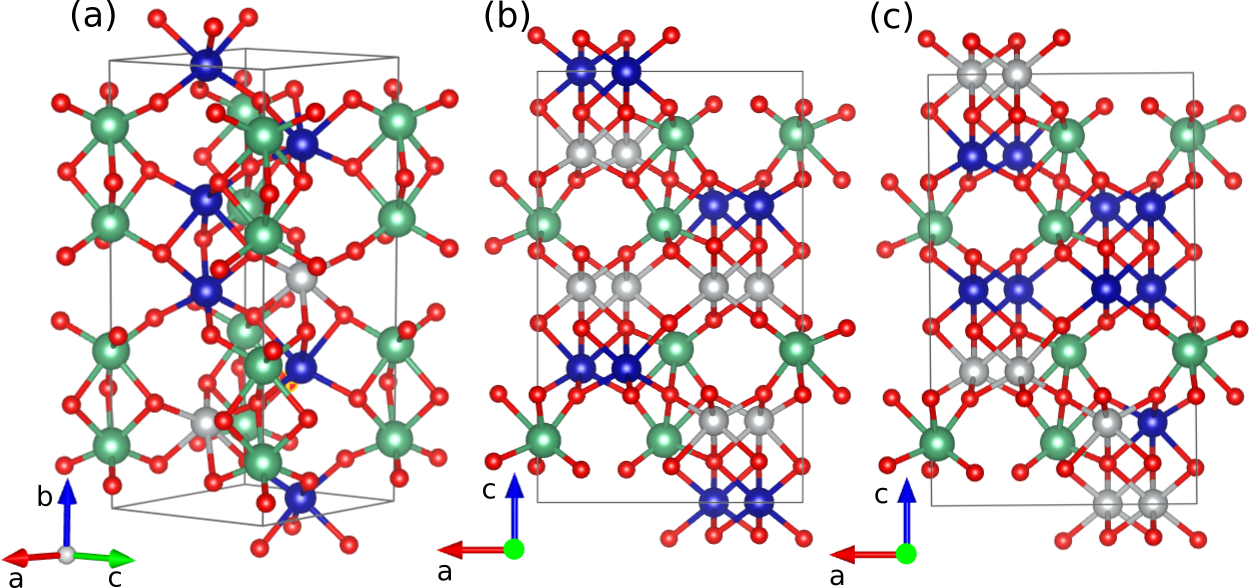}
		\caption{ The structure of the most stable combination of Co and Ni for (a) Co$_3$Ni$_1$Nb$_2$O$_9$ (CN1), (b) Co$_2$Ni$_2$Nb$_2$O$_9$ (CN2) with  0.5/0.5 occupancy of Co$_2$/Ni$_2$ and (c)  CN2 with  0.625/0.375 occupancy of Co$_2$/Ni$_2$. Co, Ni, Nb, and O are indicated 
			by blue, gray, green and red spheres, respectively.}
		\label{opt_struc}
	\end{center}   
\end{figure}

To find out which combination is energetically favorable, 
we used  spin-polarized DFT/PBE calculation by employing QE code and set the AFM configurations 
for CN1 and CN2 similar to the magnetic configuration which shown in Fig.~\ref{dmc}.
The most stable combinations of Co and Ni atoms for CN1 and CN2 (for both 0.5/0.5 and 0.625/0.375 occupancy of Co$_2$/Ni$_2$) 
are indicated in Fig.~\ref{opt_struc}.
For CN1, we selected the most stable combination among the calculations with nearly zero total magnetization.

\begin{figure}[!]
	\begin{center}
		\includegraphics[width=8.5cm,angle=0]{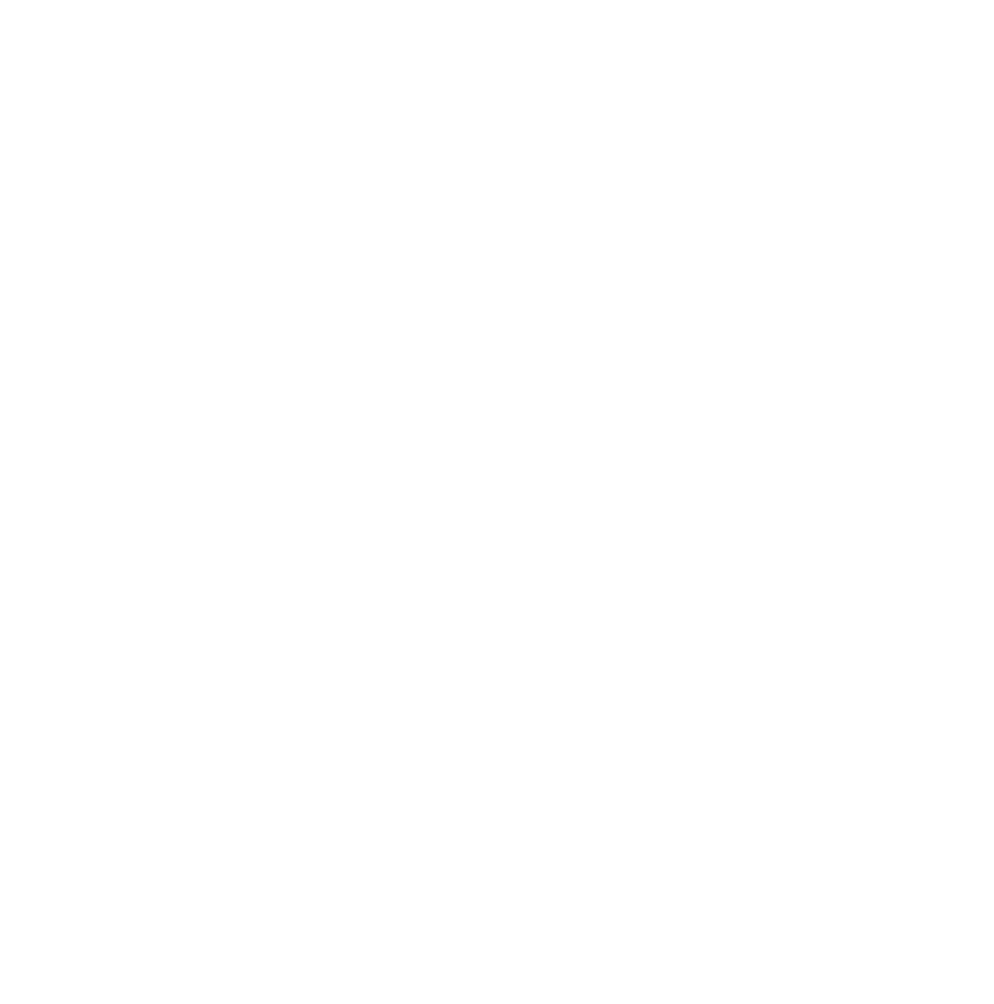}
		\caption{ The magnetic anisotropy energy (MAE) of Co$_2$Ni$_2$Nb$_2$O$_9$ (CN2) for two rotation axes \textbf{b} ($[001]\rightarrow [100]$) and  
			\textbf{c} ( $[100]\rightarrow [010]$ )  by using DFT+$U$ calculations with $U=6$ eV and $J_H=1$ eV, 
			where $U$ is Hubbard parameter and $J_H$ is Hund exchange parameter. 
			The figure (a) indicates the MAE of CN2 with 0.5/0.5 occupancy of Co$_2$/Ni$_2$ and (b) the MAE of CN2 with 0.625/0.375 occupancy of Co$_2$/Ni$_2$. The reference energy for  calculation of MAE for rotation about \textbf{a}-axis (\textbf{c}-axis) is the total  energy of CN2 with magnetic moments along \textbf{c}-axis (\textbf{a}-axis).}
		\label{mae}
	\end{center}   
\end{figure}

The easy axis of magnetic anisotropy of CN2 can determine the most probable
magnetic structure for this sample between $ \Gamma_{3} $ and $ \Gamma_{5} $ configurations.
By using Fleur/DFT+$U$ and the most stable combination of Co and Ni for 0.5/0.5 and 0.625/0.375 occupancy  of Co$_2$/Ni$_2$ 
(Fig~\ref{opt_struc}-(b),(c)),
we calculated the magnetic anisotropy energy (MAE) by rotation of magnetic moments 
around \textbf{b} (from $[001]$ to  $[100]$ direction) and \textbf{c} (from $[100]$ to $[010]$ direction) axis.
The results are shown in Fig.\ref{mae}. As can be seen, MAE decreases when the magnetic moments rotated from \textbf{c} to \textbf{a} and also from \textbf{a} to \textbf{b}, which means \textbf{b} is the easy axis. Therefore, our DFT+$U$ calculations  
suggest the $\Gamma_{5}$ magnetic structure (the one which gave slightly better agreement factors of refinements ) for CN2. 
For these calculations we set $U=6$ eV and $J_H=1$ eV (Hund exchange parameter).
To be sure that the easy axis doesn't depend on $U$ parameter, we recalculated MAE for \textbf{a}, \textbf{b} and \textbf{c} direction with $U=4$ eV.
The result indicates that \textbf{b} is still the easy axis.

\subsection{Magnetization and Susceptibility }\label{MAG}
\begin{figure}[H]
	\begin{center}
		\includegraphics[width=9cm,angle=0]{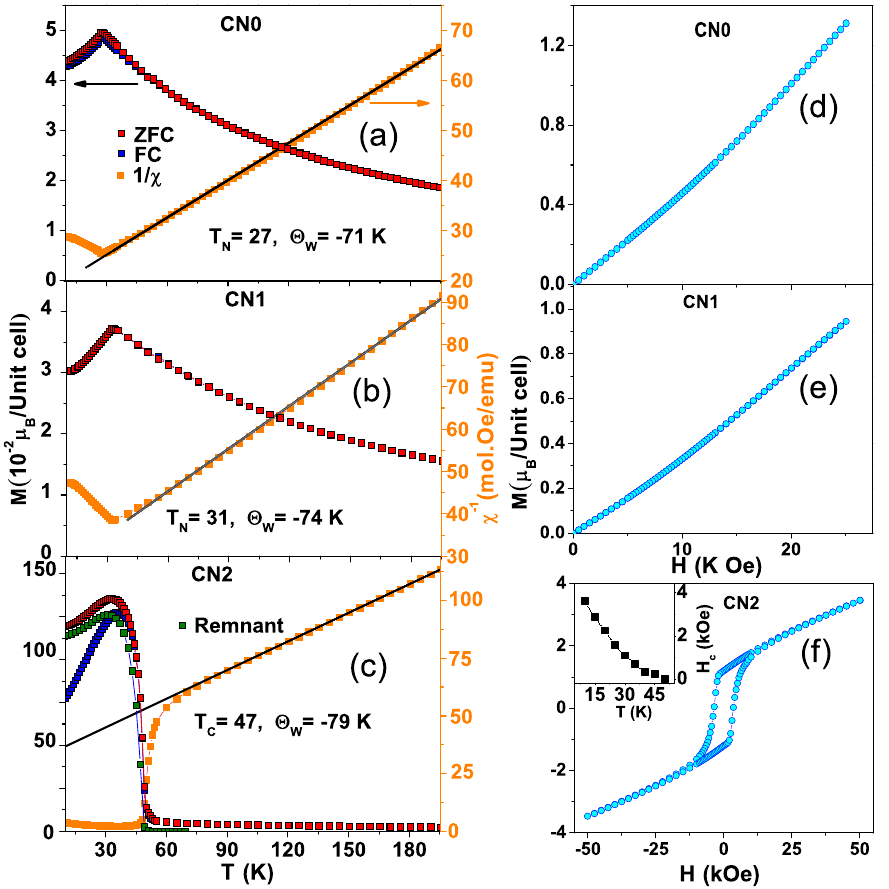}
		\caption{ ZFC and FC magnetization, inverse magnetic susceptibility and Curie-Weiss fit as functions of temperature for (a)  Co$_4$Nb$_2$O$_9$ (CN0), (b)  Co$_3$Ni$_1$Nb$_2$O$_9$ (CN1), and (c) Co$_2$Ni$_2$Nb$_2$O$_9$ (CN2); magnetization as a function of magnetic field for (d) CN0, (e) CN1, and (f) CN2; Inset of (f) shows temperature dependence of coercivity field of CN2. 
			\label{mag2}}
	\end{center}
\end{figure}

Figure \ref{mag2}(a), (b) and (c) shows the temperature dependence of FC-ZFC magnetization for CN0, CN1 and CN2, respectively. The inverse dc susceptibility, $ \chi^{-1}\textrm{(T)} $, is also included for each of the samples. An AFM transition around 27 K is observed for CN0 (Fig. \ref{mag2} (a)), which is in consistence with the earlier reports \cite{Kolodiazhnyi,Solovyev}. Fig. \ref{mag2} (b) shows that, similar to CN0, CN1 shows an AFM transition below T$_N$ around 31 K. In fact, replacing of one Co$^{2+} $ ion by Ni$^{2+} $ causes an increase in the Neel temperature and a decrease in the magnetization per unit cell. Fitting the measured $ \chi^{-1}\textrm{(T)} $ above 50 K to Curie-Weiss (CW)
law yields Weiss temperatures of $\Theta=-$71(0.1) and $ - $74(0.1) K, for CN0 and CN1, respectively. 
As shown in Fig. \ref{mag2} (c), susceptibility of CN2 increases rapidly below T$_c$=47 K, signifying a ferromagnetic-like behavior. It is also seen that the ZFC and FC curves split below 36 K.  
In order to further investigate the magnetic nature, we measured the remnant magnetization in warming run, after turning off the magnetic field at 10 K. A strong remnant magnetization was observed. $ \chi^{-1}\textrm{(T)} $ of CN2 follows CW law above 80 K with  $\Theta_{CW}=-$79(0.1) K. The occurrence of ferromagnetic-like behavior while $\Theta_{\textrm{CW}}<0$ indicates dominant AFM coupling showing a ferrimagnetic structure, in accordance with the ND results. 
Magnetization as a function of magnetic field (M(H)) measured at 10 K is plotted in Fig. \ref{mag2}(d) and (e) for CN0 and CN1, respectively. A slight change is observed in slope of the curves. Fig. \ref{mag2}(f) shows the M(H) loop for CN2 at 10 K. In contrast to the soft ferrimagnetic behavior of Ni$_4$Nb$_2$O$_9$ \cite{Ehrenberg}, this sample shows a hard ferrimagnetic behavior with a large coercivity of 3.8 kOe at 10 K. It means that the magnetic anisotropy can be controlled by Co:Ni ratio in this class of magnetoelectric materials. As shown in inset of Fig. \ref{mag2}(f), the coercivity decreases to zero when temperature reaches the transition temperature.\\

\subsection{Discussion on magnetic moments}\label{DISC}
\begin{figure}[ht]
	\begin{center}
		\includegraphics[width=8cm,angle=0]{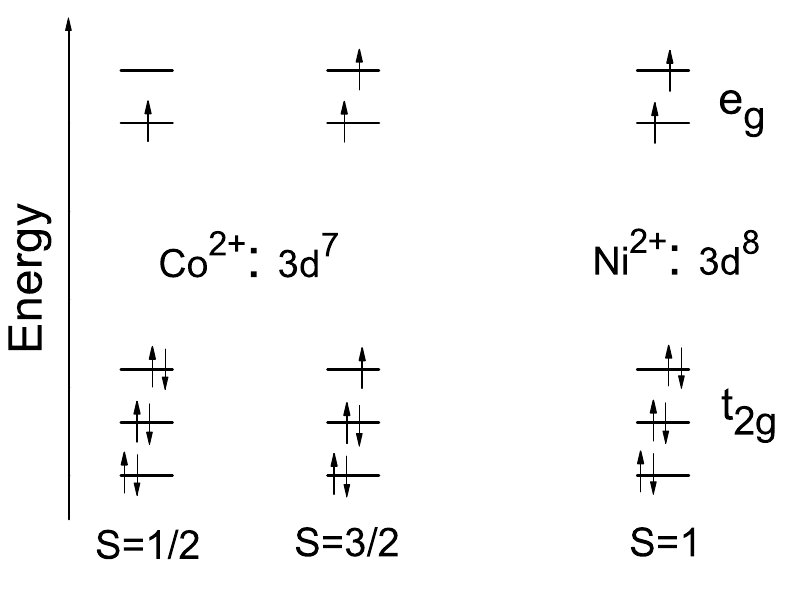}
		\caption{\label{crystalfield}Schematic of the crystal field splitting for Co$^{2+}$ and Ni$^{2+}$ ions in an octahedral environment. }
	\end{center}
\end{figure}

Here we try to interpret how cobalt and nickel contribute to the magnetic moments derived from the refinements.
Fitting the measured $ \chi^{-1}\textrm{(T)} $ above 50 K to CW
law (Fig. \ref{mag2} (a,b)) yields effective magnetic moments of $\mu_{\textrm{eff}}=$5.7 and 5.6 $\mu_{\textrm{B}}$ for CN0 and CN1, respectively. Assuming that S$_\textrm{Co}=\frac{3}{2}$ and S$_\textrm{Ni}=1$, these values are noticeably higher than those calculated for spin only ($\mu_{eff}=g\sqrt{S(S+1)}$) effective magnetic moment $\mu_{\textrm{eff}}$=3.87 and 3.77 $\mu_{\textrm{B}}$ for CN0 and CN1, respectively. Regarding the octahedral environment of magnetic ions, one can expect the t$_\textrm{2g}$-e$_\textrm{g}$ splitting due to the crystal field. As shown in Fig. \ref{crystalfield}, there are theoretically two possible spin states for Co$^{2+}$; $S=\frac{3}{2}$ and $\frac{1}{2}$ in high and low spin states. For Ni$^{2+}$ we expect $S=1$.

Having said this, we evaluate the magnetic moments derive from the refinements. When neutrons are diffracted by a magnetically ordered structure, they locally probe the magnetic moments inside the structure. Therefore, they only reveal the magnitude and direction of magnetic moment at each site but no information about the magnetic ion corresponding to the moment. Both samples studied here consist of two distinct crystallographic sites for magnetic ions and each site is shared between Co$^{2+}$ and Ni$^{2+}$ ions. Let us firstly consider the CN1 case which has similar nuclear and magnetic structures to CN0. According to Deng et al. \cite{DengPRB}, for CN0, the refined magnetic moments on the Co1 and Co2 sites are 2.32(1) and 2.52(1) $ \mu_B $, respectively. However, we would expect magnetic moments of 3 and 1 $ \mu_B $ for high spin and low spin states, respectively. Replacing a Co with Ni causes magnetic moments on both sites to decrease slightly and magnetic moments on (Co/Ni)1 and (Co/Ni)2 sites are 2.23(5) and 2.42(7) $ \mu_B $ in CN1, respectively. In other words:
\begin{equation}\label{eq9}
M1=0.75{M}_{Co1} \pm 0.25{M}_{Ni1}= 2.23 \mu_B  
\end{equation}
\begin{equation}\label{eq10}
M2=0.80{M}_{Co2} \pm 0.20{M}_{Ni2}= 2.42 \mu_B
\end{equation} 
where M1 and M2 are derived magnetic moments on site1 and site2; M$_{\textrm{Co}}$ and  M$_{\textrm{Ni}}$ are moment contributions of Co$^{2+}$ and Ni$^{2+}$ ions, respectively. "$+$" for the state that Co$^{2+}$ and Ni$^{2+}$ ions have parallel moments and "$-$" for anti-parallel state. By comparing the moment values of unsubstituted CN0 with CN1 the parallel arrangement is obviously deduced. Eqs.\ref{eq9} and \ref{eq10} lead to M$_{\textrm{Ni1}}$=1.96 and  M$_{\textrm{Ni2}}$=2.02 $ \mu_B $ for parallel state. This result is quite reasonable for Ni$^{2+}$ ion in a d$^8$ configuration with two unpaired electrons. \\The situation is more complicated for CN2 which has a different crystal structure. As it was presented in Fig. \ref{oct}, in both CN1 and CN2 cases magnetic sites are surrounded by six oxygen ions forming distorted octahedrons. Since the bond lengths of mentioned octahedrons are not different significantly, we suppose that Ni$^{2+}$ ions in CN2 structure similar to CN1 show moment of $\sim$ 2 $ \mu_B $. Thus for magnetic moments obtained from $ \Gamma_{5} $ 
\begin{equation}\label{eq11}
M1=0.50{M}_{Co2} \pm 0.50\times2= 2.44 \mu_B  
\end{equation}
\begin{equation}\label{eq12}
M2=0.60{M}_{Co1} \pm 0.40\times2=1.68 \mu_B
\end{equation}
This gives M$_{\textrm{Co1}}$=2.88 and  M$_{\textrm{Co2}}$=1.47 $ \mu_B $ for parallel state. For anti-parallel state, on the other hand, we would obtain M$_{\textrm{Co1}}$=4.13 and  M$_{\textrm{Co2}}$=6.88 $ \mu_B $. This means that at least on site2 moments of Co and Ni can not be anti-parallel because it leads to a non-reasonable value of moment for Co$^{2+}$.

\subsection{Specific heat}\label{HC}
Specific heat C$_p$ as a function of temperature and magnetic field was measured for CN0, CN1, and CN2. Fig. \ref{HC} (a) shows temperature dependent C$_p$ in zero magnetic field for CN0. In agreement with our dc susceptibility result, at 27 K, there is a $\lambda$-like peak showing the transition from a magnetically long range-ordered to paramagnetic state as evidenced by other experimental measurements discussed previously. For CN1 (Fig. \ref{HC} (b)), in zero field, a similar peak is found at 31 K. The position of this peak does not shift under a magnetic field of 140 kOe, while its magnitude drops slightly. As shown in Fig. \ref{HC} (c) for CN2, C$_p$ peaks at 47 K in zero field. Application of magnetic field shifts position of this peak to slightly higher temperatures. However, the magnitude of the peak significantly decreases with increasing field and it seems to be smeared out at high magnetic fields. These features are similar to that commonly observed at a ferromagnetic transition. Above 45 K for CN1 and 70 K for CN2, the C$_p$ for zero and 140 kOe magnetic fields converges and then increases monotonically with increasing temperature.\\ 
In order to obtain the magnetic part of the specific heat, C$_\text{mag}$, one should subtract the lattice contribution. We simulate the lattice contribution from the high temperature data by taking into account both the Debye (C$_D$) and Einstein (C$_{E}$) contributions, i.e. C$_{lattice}$=C$_D$+C$_{E}$ \cite{lin}. Details related to the method we used to analyze heat capacity data are given in the Supplementary Material \cite{supplementary}. The best fit for both zero field sets of data, using one Debye and one Einstein branch yields the characteristic temperatures and numbers: $\Theta_D$=1083(16) and $\Theta_E$=312.5(1.5) K for CN1 and $\Theta_D$=1047(65) and $\Theta_E$=322.5(10) K for CN2, $n_D$=5 and $n_E$=10 for both cases. The sum n$_D$ + n$_E$ is the total number of atoms per formula unit.\\
The extracted magnetic part of the specific heat C$_\textrm{m}$ and the corresponding magnetic entropy $S_m(T)=\int\frac{C_m(T)}{T} d T$ as a function of temperature is plotted in Fig. \ref{HC} (d-f). From $S_m=Rln(2S+1)$ for each magnetic site, a spin entropy of 10.93 and 10.33 J/mol K is theoretically expected for CN1 and CN2, respectively. The calculated S$_m$ from experimental data reaches a saturated value of 11.28 J/(mole K)  at 100 K for CN1 and 10.89 J/(mole K) at 110 K for CN2. The experimental S$_m$ is slightly higher than Rln(2S+1) for both cases, which may be due to the uncertainty of determination of lattice contribution.

\begin{figure*}[]
	\begin{center}
		\includegraphics[width=18cm,angle=0]{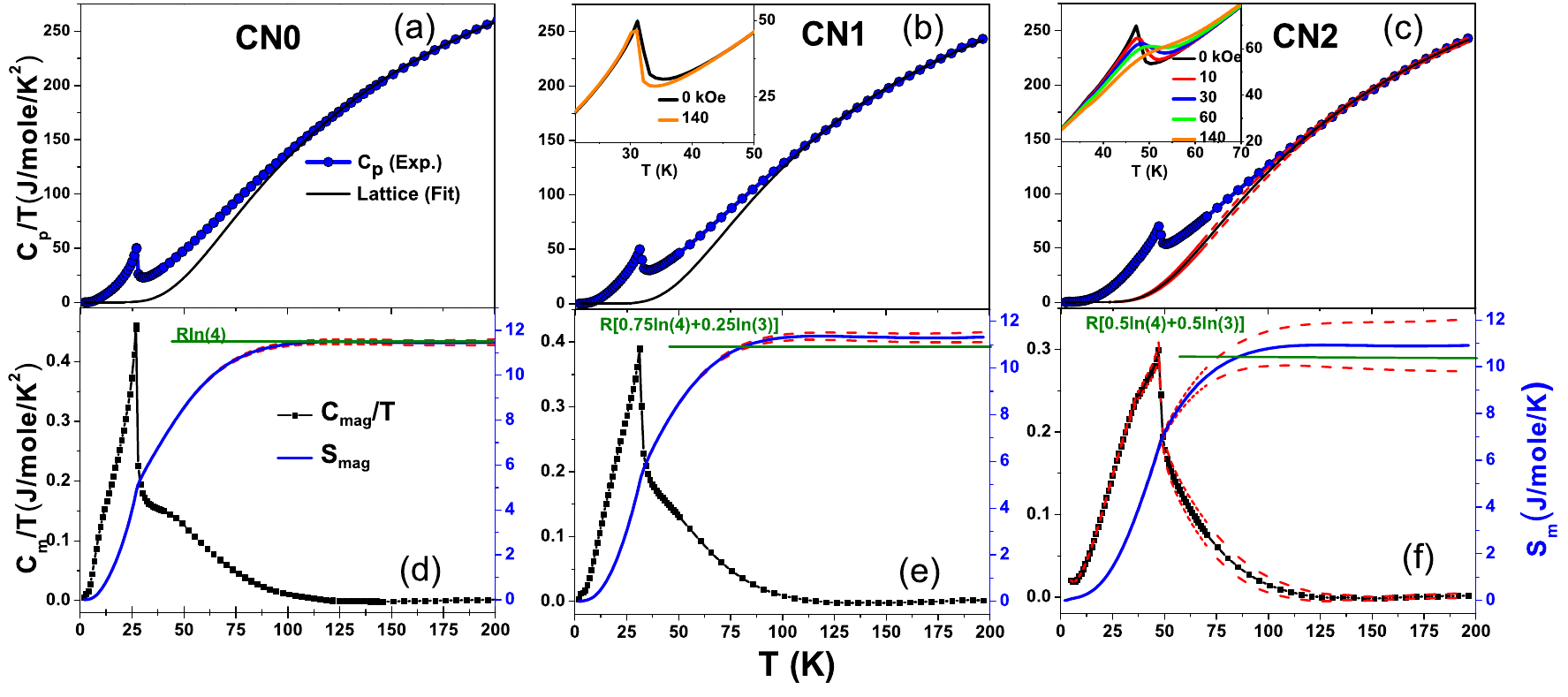}
		\caption{\label{HC}(a-c) Temperature dependence of the heat capacity C$_p$ for   Co$_4$Nb$_2$O$_9$ (CN0),  Co$_3$Ni$_1$Nb$_2$O$_9$ (CN1), and Co$_2$Ni$_2$Nb$_2$O$_9$ (CN2), respectively; The solid curve is the lattice contribution approximated by the Debye and Einstein models combined with unit cell volume (see the text). (d-f) Magnetic part of the specific heat C$_m$ divided by temperature and calculated magnetic entropy S$_m$ from C$_m$. The red dashed lines show the error bar for S$_m$ from C$_m$. Insets show the effect of magnetic field on C$_p$ of CN1 and CN2.}
	\end{center}
\end{figure*}


\section{SUMMARY}\label{SUM}
In summary, we have synthesized polycrystalline Co$_{4-x}$Ni$_x$Nb$_2$O$_9$ (x=0, 1, and 2) to investigate the role of Ni$^{2+}$ doping on the crystal structure and magnetic properties.  The compounds with $x$=0 and 1 crystallize in the trigonal P$\bar{\textrm{3}} $c1 space group. Neutron diffraction (ND) analysis reveals an in-plane weakly non-collinear AFM configuration for $x$=1. In agreement with ND results, magnetization and specific heat study show that $x$=1 undergoes an
antiferromagnetic phase transition around 31 K. On the other hand for $x$=2, the crystal structure was found to be the orthorhombic with space group Pbcn. Two possible ferrimagnetic structures with magnetic moments lying in the $ab$ plane were derived from ND data. DFT calculations distinguished the most likely magnetic configuration for this sample. The compound shows a hard- type ferrimagnetic behavior below the transition temperature of T$_c$=47 K. The heat capacity of the samples was also investigated under magnetic field up to 140 kOe. 
\section{Acknowledgement}
Work in Lausanne was supported by Swiss National Science Foundation (SNSF) grant 166298, SNSF Synergia network MPBH 160765, and European Research Council (ERC) Synergy Grant HERO.
Neutron scattering measurements were performed at the Swiss Neutron Source at Paul Scherrer Institut. This work was also partially supported by Isfahan University of Technology (IUT). 
\bibliography{references}

\end{document}